# Attosecond Field Emission

H. Y. Kim[1], M. Garg[2], S. Mandal[1], L. Seiffert[1], T. Fennel[1] & E. Goulielmakis[1]*

[1]Institut für Physik, Universität Rostock; Albert-Einstein-Straße 23–24, 18059 Rostock, Germany.
[2]Max Planck Institute for Solid State Research; Heisenbergstraße 1, 70569 Stuttgart, Germany.

*Author to whom correspondence shall be addressed: e.goulielmakis@uni-rostock.de

**Field-emission of electrons underlies major advances in science and technology, ranging from imaging the atomic-scale structure of matter to signal processing at ever-higher frequencies. The advancement of these applications to their ultimate limits of temporal resolution and frequency calls for techniques that can confine and probe the field emission on the sub-femtosecond time scale. We used intense, sub-cycle transients to induce optical field emission of electron pulses from tungsten nanotips and a weak replica of the same transient to directly probe the emission dynamics in real-time. Access into the temporal profile of the emerging electron pulses, including the duration $\tau = (53 \text{ as } \pm 5 \text{ as})$ and chirp, and the direct probing of nanoscale near-fields, open new prospects for research and applications at the interface of attosecond physics and nanooptics.**

**Main**

The interaction of atoms and molecules with intense laser fields gives rise to attosecond electron pulses[1] which can probe the structure and dynamics of these systems upon recollision with their parent ion[2]. Attosecond techniques[3,4] can now gain access to the temporal profile of the recolliding electron pulses and concomitant structural dynamics[5,6] in their parent ions by measuring the transient properties of high harmonics[7] emitted during the interaction. Studies of the interaction of intense laser fields with nanostructured metals over the last two decades have suggested that the semiclassical concepts[8–18] earlier developed to describe electron dynamics in atoms can afford a central role in the understanding of the optical field electron emission. In analogy to atoms, electrons set free from the apex of a nanotip at the field crest of an intense laser pulse should also form ultrashort electron pulses (Fig. 1a, inset), which upon recollision with the tip surface some ¾ of the laser period (T~ 2 fs) later could probe both dynamics and structure. Owing to the ultrashort time interval between emission and recollision events, and in contrast to other emerging electron pulse technologies[19–21], the electron pulse wavepacket shall undergo a negligible temporal spread allowing its confinement to the sub-cycle time scales.

Yet, the real-time tracking of electron pulses generated in the optical field emission has remained challenging. Although ordinary attosecond streaking techniques can be used to map the temporal structure of the extreme ultra-violet (EUV) electron emission from metal surfaces[22] and nanotips[23] they cannot directly probe electron pulses emerging in the optical field emission. Furthermore, the hitherto absence of high harmonic emission from laser driven nanotips constrains the applicability of in-situ attosecond techniques[24–27] for probing the structure of the electron pulses in these systems.

Harnessing electron pulses emerging in the optical field emission for realizing new spectroscopies that combine attosecond temporal and nanometer spatial resolution calls for essential advancements both on their generation and measurement methodologies. On the generation side the driving laser pulses should be both short and intense such as to confine the tunneling of electrons into a sub-femtosecond window (< 1 fs) as well as to impart the recolliding electron pulses with a de-Broglie wavelength (<2.74 Å, >20 eV) that allows atomic scale probing of the parent surface[28]. On the measurement side, in-situ attosecond measurement techniques[24–27] shall be extended to incorporate temporal gating of the optical field emission without relying on the concomitant high harmonic radiation emitted during recollision. Measurements of this kind have so far permitted access into the driving field waveform of light waves by tracking the spectrally integrated currents induced in the bulk of solids[29,30] or the cutoff energy variation of rescattered electrons in atoms[31] but a direct time-resolved measurement of attosecond electron pulses in the optical field emission has remained beyond reach.

Guided by the above provisions, we studied the field emission in tungsten nanotips (work function $\phi \sim 4.5 - 5.0$ eV) using intense ($\sim 10^{13}$ W/cm$^2$), sub-cycle (~1.9 fs) optical transients (centroid energy ~1.8 eV) generated in a light-field synthesizer[32,33]. The experiments were performed in a multifunctional experimental setup (see Methods and Extended Data Fig. 1) which combines photoemission spectroscopy of atoms and solids, optical pump-probe measurement methodologies, and EUV attosecond streaking[34,35] for the sampling of the driving field waveforms.

We commenced our experiments by probing the nonlinearity of the electron yield to the driving field intensity. We used a plate detector (see Extended Data Fig. 2) to record the total number of emitted electrons from the nanotip (yellow points, Fig. 1b) for a range of peak intensities of the impinging laser transient. Evaluation of the slope of the total electron count versus peak intensity (purple line in Fig. 1b) yielded an emission nonlinearity of $\sim 1.18 \pm 0.09$ which is well below the multiphoton threshold $\phi/\hbar\omega_L \sim 3$. This finding supports the notion that field-driven electron tunneling dominates the ionization of tungsten over the entire range of the studied intensities in these experiments.

Next, we conducted a spectral-domain study of the emission under precisely characterized driving fields and interrogated compatibility of our findings with the predictions of semiclassical, single-electron models. Given the multielectron nature of the emission (Fig. 1b) this step is essential for applying semiclassical single-electron concepts later in this work to probe the temporal structure of the electron emission. Electron spectra recorded as a function of the peak intensity of the driving pulse (Fig. 1c) revealed the formation of two, well discernible, cutoffs (black points, grey points and false color plot in Fig. 1c respectively) whose energies scale linearly with the peak intensity over the entire studied range. Evaluation of the corresponding slopes for high and low energy cutoffs ($E_c$) by linear fitting (black and grey dashed lines in Fig. 1c) of the experimental data yielded $s_{W,high}^{(exp)} = dE_c/dU_p = 118 \pm 5.1$ and $s_{W,low}^{(exp)} = dE_c/dU_p = 24.1 \pm 1.35$, respectively. Here $U_p$ is the ponderomotive energy of an electron under the driving pulse. By taking the ratio between high and low energy cutoff slopes ($\sim 4.91 \pm 0.35$) we find that it is compatible with that anticipated for backscattered and direct electrons ($10U_p/2U_p \sim 5$) [36] in the single electron approximation.

An experimental assessment of the near-field enhancement in the vicinity of the nanotip and its comparison to the theoretical predictions based on the geometrical characteristics of the nanotip, further attest to the validity of single-electron semiclassical pictures to describe emission under our experimental conditions. To this end, we compared the emission spectra of electrons from the nanotip with that in the low-density Neon gas (where field enhancement is absent) under identical driving pulses. Fig. 1d contrasts electron spectra recorded from W nanotip (red curve) and Neon atoms (blue curve) under identical field waveforms. Fig. 1e shows electron spectra emanating from Ne over a broad range of intensities of the same driving pulse. A linear fitting (Fig. 1e, dashed line) of the cutoff energy (Fig. 1e, points) of the Neon spectra versus intensity yielded a slope $s_{Ne}^{(exp)} = 9.88 \pm 0.34$ that agrees very well with the semiclassical predictions for rescattered electrons in atoms for our driving transient waveforms ($s_{Ne} = 10.8$) (see Methods). Based on the atomic measurements, the field enhancement factor in the vicinity of the tungsten nanotip is evaluated as $f = \sqrt{\frac{s_{W,high}^{(exp)}}{s_{Ne}^{(exp)}}} = 3.46 \pm 0.10$ and shows a fair agreement with the theoretical prediction, $f_{th} \sim 3.8$ (Methods).

This comparison further attests to the compatibility of the emission processes from the nanotip with single electron, semiclassical predictions.

Single and multielectron semiclassical simulations based on the experimentally derived quantities (see Methods) further verify the above perspective. The simulations accurately reproduce the experimental spectra (Extended Data Fig. 3) and the association of high and low-energy cutoffs of the emission to backscattered and direct electrons, respectively, while the inclusion of multielectron interactions (Extended Data Fig. 5) showed marginal effects of space-charge on the emitted photoelectron spectra. Moreover, the absence of discernible cutoffs at intermediate energies (60-120 eV) in both experiments (Fig. 1c) and simulations (Extended Data Figs. 3 and 5) supports the notion that under sub-cycle driving, the high-energy part of the emitted spectrum is associated with the recollision of a single electron pulse at the tip surface. A weak, low-energy backscattered emission suggested by the simulations (Extended Data Fig. 3b) is not directly resolved in the experiments (Fig. 1c). Yet the presence of such emission channels will become apparent later in this work by time-resolving the optical field emission.

**Homochromic Attosecond Streaking**

The waveform of an electron pulse generated by a laser field evolves as it experiences the forces of the oscillating electric field ensuing its birth. Therefore, a temporal characterization of the electron wavepacket has a concrete meaning at a specific position in space during its propagation. In optical field emission, the generated electron pulses probe the "sample" during the recollision with the parent surface. Thus, access into the temporal structure of the electron pulse at the surface is mostly relevant for harnessing the power of these pulses in time-resolved applications.

To understand how, we revisit the process of strong field recollision of an electron wavepacket under an intense optical waveform (Fig. 2). Set free around the peak of a laser field crest an electron pulse will recollide with the tip surface at an instance $t_r$ (Fig. 2a) with an energy of $\sim 3U_p$ [2,37]. An attosecond measurement of the electron pulse entails access into its waveform $\psi_r(t)$ at the surface of the nanotip, or equivalently, into its associated complex spectral amplitude $\tilde{\psi}_r(p)$, where $p$ is the electrons recollision momentum. However, as $\tilde{\psi}_r(p)$ is not directly accessible in measurements it is important to link it to other measurable quantities. Following back-scattering off the metal surface, the wavepacket acquires additional phase both from the interaction with the driving field (Volkov phase) as well as due to its free-space propagation. If we define an auxiliary terminal wavepacket $\tilde{\psi}_t(p)$ that includes the Volkov phase imparted to the electron wavepacket within the remaining part of the laser pulse, the spectral intensity $I(p) = |\tilde{\psi}_t(p)|^2$ (as marked in Fig. 2a)[38–41] can be written as:

$$I(p) = |\tilde{\psi}_t(p)|^2 = \left| \int_{-\infty}^{\infty} \psi_r(t_r) \exp\left[i\frac{p^2}{2} t_r\right] \exp\left[-iS\left(p, t_r; A_p(t)\right)\right] dt_r \right|^2 \quad (1)$$

where $S\left(p, t_r; A_p(t)\right) = \int_{t_r}^{\infty} \left[\frac{1}{2}[p + A_p(t)]^2 - \frac{1}{2}p^2\right] dt$ denotes the Volkov phase imparted to the electron wavepacket by the vector potential $A_p(t)$ of the intense driving pulse (hereafter referred to as pump, cf. red curve in Fig. 2a) after rescattering at time $t_r$. Hence, reconstruction of the recolliding wavepacket $\psi_r(t)$ should be possible, if other than $I(p)$, which is a directly measurable quantity (i.e, the spectrum of the electron emission), the phase of $\tilde{\psi}_t(p)$ as well as $A_p(t)$ could be accessed.

Access to the phase of $\tilde{\psi}_t(p)$ is possible by temporally gating the electron emission with a weak replica of the pump pulse (hereafter referred to as gate with vector potential $A_g(t)$) when $\eta \equiv |A_g(t)|^2 / |A_p(t)|^2 \ll 1$ [27,31]. In this limit, the pump pulse is solely responsible for releasing the electron wavepacket, while the gate pulse primarily alters its phase. This is manifested by the shift and reshaping of the terminal photoelectron spectra at the end of the driving pulse (Fig. 2a). If the delay $\tau$ between pump and gate pulses is varied (see Methods), the terminal spectral distribution of the released electron, $I(p,\tau) = |\tilde{\psi}_t(p,\tau)|^2$ can be approximated as:

$$I(p,\tau) \propto \left| \int_{-\infty}^{\infty} \psi_t(t_r) \exp\left[i\frac{p^2}{2}t_r\right] \exp[-iS(p,t_r;A_{HAS}(t))] \, dt_r \right|^2 \qquad (2)$$

where $\psi_t(t)$ is the inverse Fourier transform of $\tilde{\psi}_t(p)$, and $A_{HAS}(t+\tau)$ represents an effective vector potential, which is explicitly related to the incident vector potential $A_g(t)$ of the gate pulse as shown in Methods and which accounts for the momentum an electron accumulates from the instance of its birth to the detection. Eq. 2 implies: (i) A variation of $\tau$ permits the composition of a streaking-like spectrogram whose reconstruction can retrieve the phase of $\tilde{\psi}_t(p)$. (ii) The momentum variation of the electron distribution follows $A_{HAS}(t)$. Whereas implication of (ii) (see Methods) allows sampling of the waveform of a light pulse[31], (i) is essential for mapping the dynamics of the field emission. To distinguish from conventional attosecond streaking in this work, we refer to this approach as *Homochromic Attosecond Streaking* (HAS), highlighting that the carrier frequency of pump and gate fields is identical.

Shown in Fig. 2b are simulated HAS spectrograms under conditions pertinent to the experiments presented later in this work. Notably, and in close analogy to ordinary attosecond streaking[42,43], different types of chirps of the recolliding electron pulse $\psi_r(t)$ yield distinct visual signatures in the spectrogram, manifested as shifts of the intensity modulation of the spectra versus delay and energy as highlighted by the white dashed-curves in Fig. 2b.

To utilize the principles of HAS in our experiments, we derived the pump and gate pulses by the spatial and temporal division of the sub-cycle optical transients using the dual mirror module shown in Fig. 1a. Fig. 3a shows a HAS spectrogram recorded by our apparatus. We verified that the gate pulse is sufficiently weak to serve as a phase gate (Eq. 2) by conducting a systematic study presented in Methods, where we compared the vector potential of the gate pulses as evaluated from HAS and conventional EUV-based attosecond streaking spectrograms for gradually increasing ratio of $\eta = |A_g(t)|^2/|A_p(t)|^2$. Our experiments suggest that a ratio below $10^{-2}$ is adequate to reliably measure the electric field waveform. A remaining weak amplitude modulation of the spectrogram of the order of 5-10% turns out to be useful for evaluating the absolute delay between pump and gate pulses, and therewith to clock the recollision instance with respect to the waveform of the driving pump pulse without resorting to additional measurements.

For the case study of the data of Fig. 3a (ratio $\eta \sim 6.3 \times 10^{-3}$) $A_g(t)$ (blue curve in Fig. 3b) exhibits an excellent waveform-matching with that evaluated by EUV attosecond streaking (red curve, Fig. 3b) attested to by a degree of similarity[44] of ~0.95. In this case the HAS-based evaluated vector potential clearly represents the near-field of the gate pulse in the vicinity of the tip. Thus, the ratio between the absolute amplitudes of the vector potentials evaluated with the two methods provides a direct, time-domain measurement of the field enhancement factor $f = 3.74 \pm 0.25$, which is in close agreement with the result of the methodology of Fig. 1c (conducted with a different nanotip) and the theoretical estimations.

How time-domain measurement of electron emission phenomena in nanostructured materials benefits the intuitive understanding of the processes can be best appreciated by a close inspection of the oscillating phases of high and low energy cutoffs versus delay in Fig. 3a. For instance, an apparent delay of the maxima of the corresponding oscillations (highlighted by red and blue dashed curves respectively in Fig. 3a) versus delay indicates a retardation in the electron emission at lower energies by approximately a laser-cycle ($T_L \sim 2.3$ fs). This feature, also well reproduced in our simulation (Fig. 3c), reveals that the low energy emission is comprised by a mixture of direct electrons emerging within the main half cycle of the driving field (green dots in Fig. 3d) and low-energy backscattered electrons generated approximately a cycle of the driving field later (purple dots, Fig. 3d). Thus, HAS offers direct, real-time clocking of the optical field emission. More importantly, the uniform amplitude and energy modulation of the photoelectron spectrum, (Fig. 3a) over a broad range of energies (50-150 eV), offers evidence that a single electron pulse confined to a fraction of a field half-cycle is responsible for the emission.

The above conclusions allow us to now turn to the most central aspect of this work which is the reconstruction of the temporal profile of the recolliding electron wavepacket. We are primarily interested in the properties of the recolliding electron pulses at energies typically higher than 20 eV, i.e., when this pulse could serve future high-resolution/atomic-scale, spatial probing of surfaces[28]. Considering the ratio of the terminal energy of electrons ($\sim 10 U_p$) and that at the recollision instance ($\sim 3.17 U_p$) the relevant temporal information associated with the recolliding electron pulses shall be sought at the high-energy end ($> 80$ eV) of the spectrogram of Fig. 3a.

Figure 4a isolates the relevant section of the spectrogram derived from Fig. 3a (>80 eV), while Fig. 4b shows its numerical reconstruction based on Eqs. 1 and 2 with the retrieved field parameters $A_p(t)$, $A_g(t)$ (Fig. 3b), absolute time delay $\tau$ (Extended Data Fig. 7) and the numerical algorithm detailed in Methods.

The retrieved electron pulse profile in spectral and time domains are shown in Fig. 4c and 4d, respectively. The pulse spectrum extends over the energy range 20-80 eV (Fig. 4c, magenta shaded curve), and is temporally confined to ~53±5 attoseconds as measured at the full width at the half-maximum of the retrieved intensity profile (Fig. 4d magenta shaded curve). Fig. 4e juxtaposes the electron pulse and the near-field profiles and suggests that the electron pulse revisits the surface of the nanotip (magenta shaded curve) at times close to the zero transition of the driving field (red-curve). This observation is compatible with the semiclassical understanding of strong field rescattering in atoms[2,36].

A closer inspection of the retrieved spectral and temporal phases (red curves Fig. 4c and Fig. 4d) reveals a negligible temporal spread of the electron pulse compared to its Fourier limited duration (~50 as). The compatibility of this finding with the semiclassical model can be best appreciated by comparing the group delay of the electron emission (black solid line in Fig. 4f) as evaluated by the time-frequency analysis (Fig. 4f, false color) of the attosecond electron pulse waveform in Fig. 4d with the semiclassical emission times (red dash curve, Fig. 4f) calculated using the near-field waveform of the driving pulse (red curve, Fig. 4e). The comparison highlights that, as anticipated by the semiclassical recollision model, electrons of energy close to the cutoff, i.e. where short and long trajectories merge, carry a minor temporal chirp. This feature which has been often observed in near cutoff high harmonic emission in atoms[24,42] further strengthens the link of strong-field optics in atoms and nanooptics systems.

**Conclusion**

Direct measurement of attosecond electron pulses in the optical field emission significantly extends the repertoire of ultrafast science tools that allow access to ever more complex systems. Application of HAS to systems rich to plasmonic resonances could allow the direct, real-time measurement of ultrafast collective electronic phenomena on nanometer scale dimensions. The energetic recolliding electron pulses shall also enable attosecond-resolved electron diffraction experiments at nanotips as well as nanodiffraction experiments on individual molecules attached on the nanotips. Such possibilities, which are now becoming available in molecular gas phase systems under strong fields[1,45], could be ported to solids enabling new routes for exploring the structure of condensed matter in four dimensions.

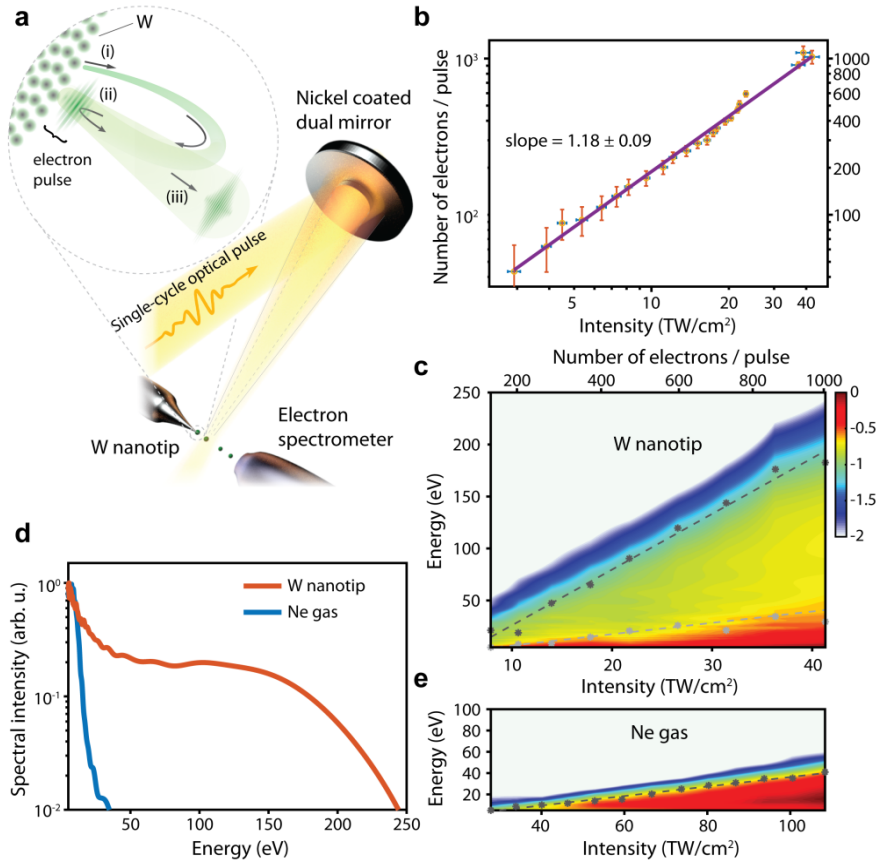

**Fig. 1 | Optical field emission by intense, sub-cycle optical transients. a**, Simplified schematic of the experimental setup. A sub-cycle pulse (orange curve) is spatially separated and focused by a dual concave Ni mirror module. A time delay between the pulses reflected off by the inner and outer mirrors is introduced by a piezo stage. Tungsten nanotips (apex radius ~35 nm) or a gas jet of Neon atoms can be positioned in the laser focus. Emitted electron spectra are recorded by a time-of-flight spectrometer (TOF) (acceptance angle ~6°) placed ~3 mm downstream the electron source and aligned along the laser polarization axis. Inset shows electrons marked by the green shaded curve are (i) set free and are accelerated by the intense laser field to form an electron pulse which up recollision with the nanotip surface (ii) can probe both dynamics as well as structure. Upon backscattering off the tip surface, the electron pulse is further accelerated by the laser to escape the interaction area (iii). **b**, Total electron yield per pulse as a function of increasing peak intensity of the driving laser pulses (yellow points) and its linear fitting (purple line) on a logarithmic scale. **c**, Electron spectra

from the tungsten nanotip (false color) versus peak intensity. Stars and dots denote the cutoff energies. Black and grey dashed lines show the linear fitting of the cutoff energy versus incident peak intensity of the laser pulse. **d**, Optical emission electron spectra from the tungsten nanotip (red curve) and neon atoms (blue curve) for nearly identical peak intensity (∼40 TW/cm$^2$). **e,** Same as in (**c**) but for neon atoms.

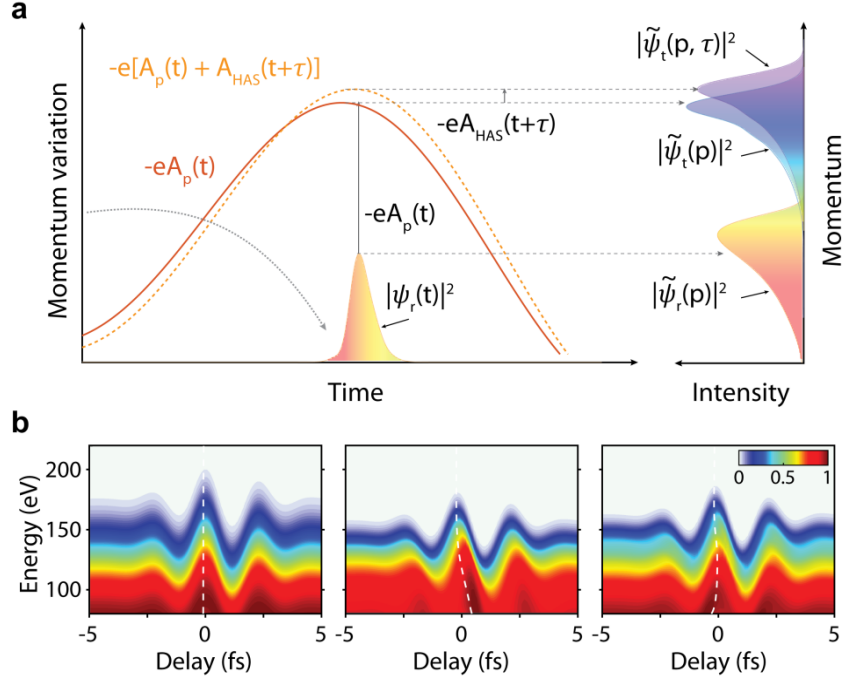

**Fig. 2 | Homochromic Attosecond Streaking (HAS). a**, An electron wavepacket $\psi_r(t)$, sampled upon recollision at the nanotip surface, is represented by its momentum distribution $\tilde{\psi}_r(p)$ and corresponding spectral intensity $|\tilde{\psi}_r(p)|^2$. Following backscattering further acceleration of the wavepacket by the vector potential of the pump field $A_p(t)$ (red curve) gives rise to a terminal wavepacket $\tilde{\psi}_t(p)$ with corresponding spectral intensity $I(p) = |\tilde{\psi}_t(p)|^2$. The coherent superposition (yellow dashed curve) of the pump field with a weak, time delayed ($\tau$) replica (the gate) results in a momentum shift, $-eA_{HAS}(t+\tau)$ of the final momentum distribution $I(p,\tau) = |\tilde{\psi}_t(p,\tau)|^2$. **b,** Simulated delay-dependent HAS spectrograms for electron pulses without chirp (left panel), positive chirp of $3.5 \times 10^3$ as$^2$ (middle panel) and third order chirp of $1 \times 10^5$ as$^3$ (right panel). White dashed curves highlight the resulting energy-dependent shifts of the intensity modulation.

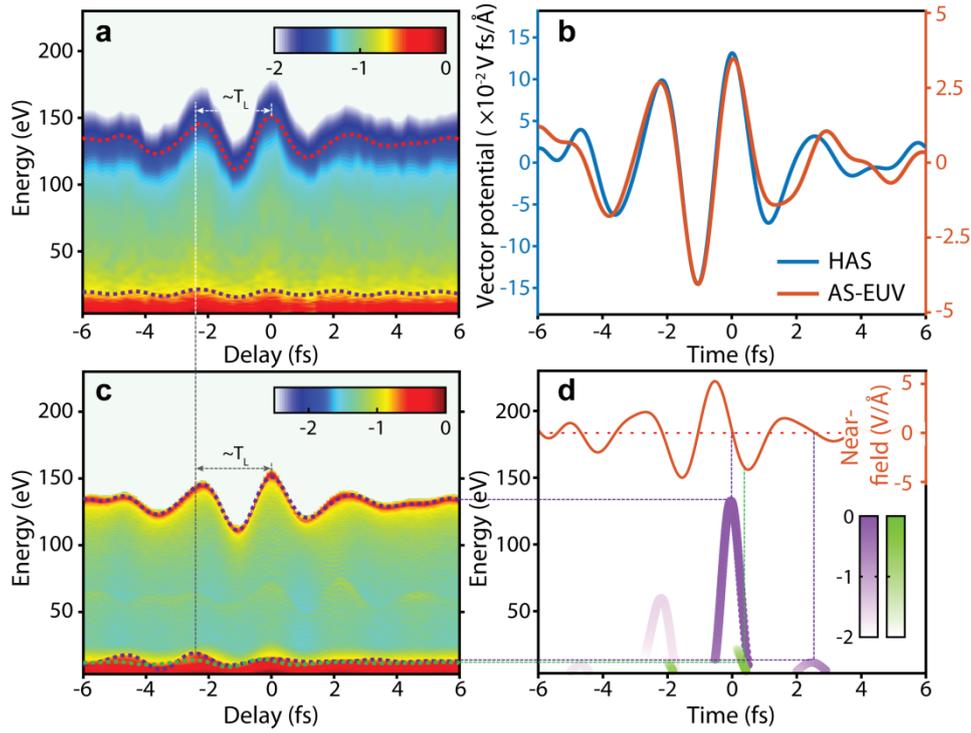

**Fig. 3 | Homochromic attosecond streaking of the optical field-emission. a**, HAS spectrogram of the optical field emission from a tungsten nanotip comprised by 60 individual electron spectra recorded as a function of the delay (steps size of 200 as) between an intense ($I_p \approx 25.4$ TW/cm$^2$) single-cycle pulse and a $6.3 \times 10^{-3}$ times weaker gate pulse. **b**, Vector potential $A_g(t)$ evaluated by the photoelectron energy cutoff variation in the HAS spectrogram in panel **a** (blue) as a function of the delay and (red) from an ordinary streaking spectrogram (see Extended Data Fig. 8) in absolute units. **c**, Simulated HAS spectrogram using the experimentally recorded field waveforms of pump and gate pulses. The vertical dashed lines in (**a**) and (**c**) denote different rescattering events. **d**, Terminal energies of direct (green) and backscattered electron emissions (purple) associated (horizontal dashed lines) with corresponding cutoff energies in the spectrogram (**c**) as well as release times (vertical dashed lines) near-field pump waveform (red curve).

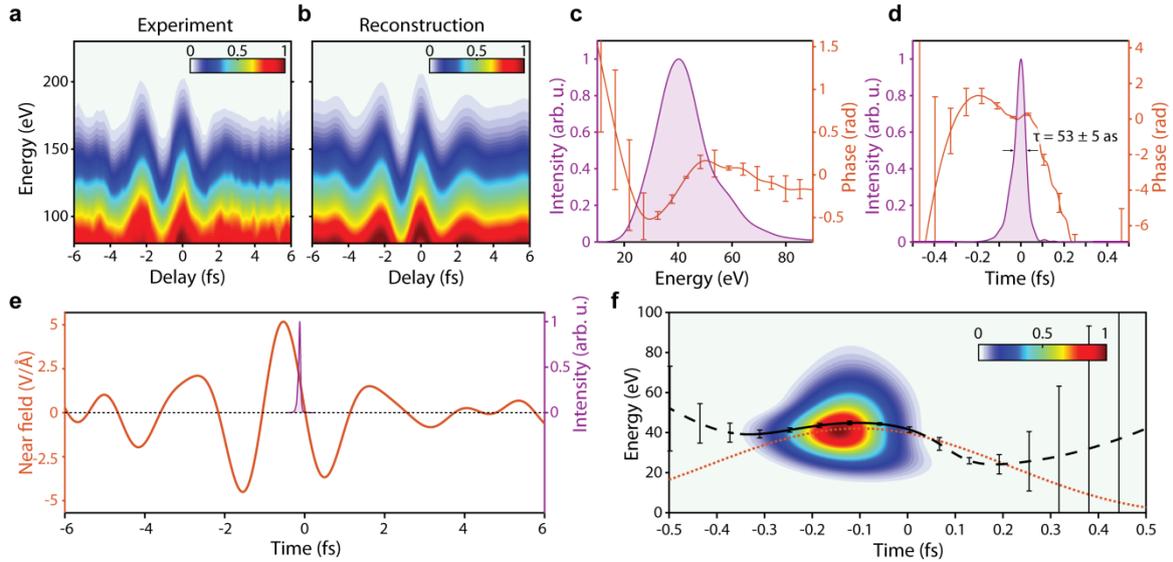

**Fig. 4 | Measurement of attosecond electron pulses in the optical field emission. a, b**, Experimentally recorded (**a**) and reconstructed (**b**) HAS spectrograms from a W nanotip. **c**, Retrieved backscattered electron pulse spectrum (magenta fill) and its spectral phase (red curve). **d**, Intensity profile of the electron pulse (magenta fill) and temporal phase (red line). **e**, Near-field (red curve) and its timing with respect to the attosecond electron pulse (magenta fill). **f**, Time-frequency analysis of the attosecond electron pulse (false color is intensity in arbitrary units) and retrieved emission times (black curve). The red dashed curve denotes emission times calculated semi-classically for the near-field light waveform in (**e**). Error bars represent standard deviations of the corresponding values indicated.

# Methods

**Experimental**

**Attosecond EUV streaking.** For the attosecond EUV streaking measurements (Extended Data Fig. 1a), the sub-cycle transients are focused onto the neon gas jet to generate extreme ultra-violet (EUV) pulses by high-harmonic generation. The colinearly propagating EUV and optical pulses are spatially separated by a Zr-filter which also acts as an EUV high pass spectral filter (> 70 eV) enabling the isolation of a single attosecond pulse. The EUV and optical pulses are reflected off a dual mirror assembly which consists of a Mo/Si inner mirror (centered at ~85 eV) and a nickel outer mirror, respectively. Inner and outer mirrors can be delayed with nanometric resolution (Extended Data Fig. 1a). EUV and optical pulses are focused onto a second Ne gas jet. Photoelectron spectra recorded as a function of the delay between the inner and outer mirrors allow the composition of attosecond streaking spectrograms which allow the detailed characterization of the attosecond EUV pulse, and more importantly for these experiments the field waveform of the optical pulse. Details about the relevant techniques can be found in references[34,35,42].

**Homochromic Attosecond Streaking.** Homochromic Attosecond Streaking (HAS) measurements are performed on the same setup via an (a) automatized removal of the Ne gas used to generate high harmonics and the Zr-filter, (b) automatized replacement of the inner mirror in the dual mirror module by a Ni-coated one of the same focal lengths (Extended Data Fig. 1b) and (c) the streaking gas nozzle is replaced by a tungsten nanotip. The above setup modifications are executed in a fraction of a minute and warrant identical conditions for all relevant measurements. In the HAS configuration of the setup in Extended Data Fig. 1b, Ni-coated inner and outer mirrors spatially divide the optical pulse into pump (inner mirror beam) and gate (outer mirror) pulses. A delay between the pulses reflected off inner and outer mirrors respectively is introduced by a precision transitional stage (see inset in Extended Data Fig. 1b).

**Measurement of the absolute electron yield in the optical field emission.** For the measurement of the total electron counts per pulse generated in our setup a thin electrode (size of ~ 5 mm x 5 mm) is introduced ~ 2 mm above the nanotip (Extended Data Fig. 2). This configuration allows detection of released electrons over a solid angle $\Omega > \pi$ steradians. The induced voltage on the thin plate is measured by a lock-in amplifier at the reference frequency of the repetition rate of the driving laser (~3 KHz). The electronic current is evaluated by dividing the induced voltage by the system impedance (10 MΩ). The total electron count per pulse is in turn obtained by dividing the current by the repetition rate of the laser and the electon charge.

**One dimensional, semiclassical simulations of the optical field emission**

The time dependent ionization probability from a tungsten nanotip was approximated by the Fowler-Nordheim formula as[12,17,46,47]:

$$p(E(t)) = N\theta(-E(t))|E(t)|^2 \exp\left[-\frac{4\sqrt{2m}\phi^{3/2}}{3\hbar e|E(t)|}\right], \qquad (3)$$

where $E(t)$ is the electric field waveform of the driving pulses, $\phi$ is the workfunction of the metal and $m$, $\hbar$ and $e$ are electron mass, reduced Planck's constant and electron charge, respectively. We calculated electron trajectories using the classical equations of motion in the single-electron limit as[2,15,17,18,36]:

$$m\frac{d^2 z_i}{dt^2} = -ef_0 E(t) \qquad (4)$$

Here, $i$ is the index of each individual trajectory and $f_0$ is the field enhancement factor. At the end of the pulse, an electron spectrum is evaluated by a spectral binning of the energies of all trajectories weighted by the ionization rate at the instances of their births.

For the experiments described here we simulated electron spectra from tungsten ($\phi = 4.53$ eV) versus peak field intensity of the driving pulse (Extended Data Fig. 3). The driving field (red, Extended Data Fig. 3a) used in our simulations was measured by an EUV attosecond streaking setup[34,35] integrated in our apparatus. A field enhancement factor of $f_0 = 3.46$ used in these simulations was derived experimentally as described in the main text.

In agreement with the data of Fig. 1c, the simulated electron spectra exhibit two well-discernible energy cutoffs, red and blue dashed lines, (Extended Data Fig. 3b) associated with the back-scattered (purple line in Extended Data Fig. 3a) and the direct (green line in Extended Data Fig. 3a) electrons. The slopes of high ($s_{W,high}^{(th)} = dE_c/dU_p = 130.3$) and low ($s_{W,low}^{(th)} = 26.1$) energy cutoffs agree well with those in our measurements (Fig. 1c). The theory reveals additional emission cutoffs at energies lower than that of the direct electrons. Because these are relatively weak, they do not leave any direct signatures in the photoelectron spectra. Yet such contributions become visible in HAS spectrograms as discussed in Fig. 3 of the main text.

**FDTD simulations of the field enhancement**

To theoretically estimate the nearfield enhancement in the vicinity of the tungsten nanotip we numerically solved Maxwell's equations via three-dimensional finite-difference time domain (FDTD) simulations. The nanotip was modeled as shown in Extended Data Fig. 4a as a sphere with radius of 35 nm that smoothly transitions to a cone with an opening angle (single side) of 15 ° and considering optical properties for tungsten[48]. The simulations predict a peak field enhancement factor of ~3.8 close to the surface at the tip apex. For comparison, the spatial distribution of the enhancement at a respective tungsten nanosphere (i.e. excluding the cone) is shown in Extended Data Fig. 4b, revealing a slightly lower peak enhancement factor of ~2.7.

## Three-dimensional semiclassical trajectory simulation including charge interaction

To inspect whether charge interaction significantly impacts the electron emission dynamics for the considered parameters, we performed semiclassical trajectory simulations utilizing the Mie-Meanfield-Monte-Carlo (M³C) model[49]. The latter has been used extensively for the study of strong-field ionization in dielectric nanospheres[50–52] and has recently been adopted also for the description of metallic nanotips[53]. In brief, we mimic the apex of the nanotip as one half of a sphere with corresponding radius. The nearfield is evaluated as the combined linear polarization field due to the incident pulse (evaluated via the Mie solution of Maxwell's equations) and an additional nonlinear contribution due to charge interaction treated as a mean-field in electrostatic approximation (evaluated by high-order multipole expansion). The latter includes Coulomb interactions among the emitted electrons and as well as their image charges (i.e. an additional sphere polarization caused by the free electrons). Photoelectron trajectories are generated by Monte-Carlo sampling of ionization events at the sphere surface, where we evaluate tunneling probabilities within WKB approximation by integration through the barrier provided by the local nearfield. Trajectories are propagated in the nearfield by integration of classical equations of motion and accounting for electron-atom collisions via respective scattering cross-sections for electrons moving within the material. To mimic the slightly higher peak enhancement of the linear response nearfield at a tungsten tip ($\approx 3.8$) as compared to a sphere ($\approx 2.7$), see Extended Data Fig. 4, we rescaled the incident laser intensity by a factor of 1.4. The so performed M³C simulations predict about three times less emitted electrons than the experiments, which we attribute to contributions of slow electrons originating from the shank of the nanotip. This is substantiated by comparison of the total electron yields predicted for nanotip and half sphere, obtained via integration of the local ionization rates over the respective surface areas and the pulse duration. However, the charge densities at the pole of the sphere and the tip apex are comparable, enabling to inspect the impact of charge interactions within the simplified simulation model. In the considered range of laser intensities, the modification of the spectral shape and, most importantly, the cutoff energy of rescattered electrons is negligible when charge interactions are taken into account (compare spectra in Extended Data Figs. 5a and b), underpinning the applicability of the proposed HAS metrology.

## Mathematical formulation of Homochromic Attosecond Streaking

Key objective of HAS is to retrieve the temporal structure of an attosecond electron pulse wavepacket $\psi_r(t)$ at the recollision on its parent surface. As this wavefunction is not directly accessible, it is necessary to link it to other quantities that are either directly measured in the experiments (such as the terminal spectral intensity $I(p) = |\tilde{\psi}_t(p)|^2$ at a detector) or can be reconstructed from the experimental data.

**Description of strong-field electron emission.** Considering an electron released from and driven back to a surface by a strong pump field $E_p(t)$, its recolliding wavepacket $\psi_r(t)$ can be linked to its terminal spectral amplitude $\tilde{\psi}_t(p)$ at the end of interaction with the driving pulse. Thereto, the time-dependent recollision wavepacket $\psi_r(t)$ is expressed via its Fourier representation $\tilde{\psi}_r(p) = \int_{-\infty}^{\infty} \psi_r(t_r) \exp\left[i\frac{p^2}{2}t_r\right] dt_r$ and following the recollision the spectral amplitude is transformed to the terminal form[38–41]:

$$\tilde{\psi}_t(p) \propto i \int_{-\infty}^{\infty} \psi_r(t_r) \exp\left[i\frac{p^2}{2}t_r\right] \exp[-iS(p, \infty, t_r; A_p(t))] \, dt_r \tag{5}$$

Here, $S$ is the Volkov phase imparted to the electron wavepacket only by the vector potential $A_p(t)$ of the pump pulse after recollision at an instance $t_r$ where the general form of the Volkov phase accumulated from a time instance $t_1$ to a later instance $t_2$ by an electric field with vector potential $A(t)$ is expressed as[54]:

$$S(p, t_2, t_1; A(t)) = \int_{t_1}^{t_2} \left[\frac{1}{2}[p + A(t)]^2 - \frac{1}{2}p^2\right] dt \qquad (6)$$

Note that Eq. 6 excludes free space propagation, i.e., it vanishes in the absence of the field and Eq. 5 reflects the momentum-dependent wavefunction at the surface including phases accumulated only by the pump field.

Earlier semiclassical theories of strong field emission[39–41] in atoms have suggested that the recolliding wavepacket $\psi_r(t)$ can be expressed by integration over ionization instance $t'$ prior to recollision at time $t_r$ and over canonical momenta $p'$ in terms of the ionization amplitude, dictated by the dipole transition $E_p(t')\, d(p' + A_p(t))$, the scattering amplitude typically described as $g(p' + A_p(t_r))$ and the Volkov phase that the electron accumulates from $t'$ to $t_r$ as:

$$\psi_r(t_r) = \int_{-\infty}^{t_r} \int g\left(p' + A_p(t_r)\right) E_p(t') d\left(p' + A_p(t')\right)$$
$$\times \exp[-iS\left(p', t_r, t'; A_p(t)\right) - i\frac{p^2}{2}(t_r - t') + i\phi t']\, dp'\, dt' \qquad (7)$$

Here $d$ and $g$ are the dipole and scattering matrix element, respectively as defined in refs.[39–41] and $e^{i\phi t}$ reflects the additional phases acquired during the time evolution of the bound state prior to ionization.

**Description of electron wavepackets under addition of a weak gate field.** Equation 5 implies that access to $\psi_r(t)$ is possible if $\tilde{\psi}_t(p)$ and $A_p(t)$ are known. Therefore, our goal is to describe how these quantities can be accessed using a phase gating process of the optical field emission by a weak replica of the driving pulse (gate pulse). Now we inspect the effects of adding the gate pulse on the dynamics of the electron described in Eqs. 5-7. We define the gate pulse by a field $E_g(t+\tau)$ and its vector potential $A_g(t+\tau)$, where $\tau$ is the delay between the pump and gate pulses as described above. By replacing the pump fields and its vector potentials by the superposition of pump and gate pulses in Eqs. 5-7 as $E_p(t) \to E_p(t) + E_g(t+\tau)$ and $A_p(t) \to A_p(t) + A_g(t+\tau)$, the terminal spectral amplitude perturbed by the gate can be rewritten as:

$$\tilde{\psi}_t(p,\tau) \propto i \int_{-\infty}^{\infty} \psi_r^{(g)}(t_r,\tau) \exp\left[i\frac{p^2}{2}t_r\right] \exp[-iS(p,\infty,t_r; A_p(t) + A_g(t+\tau))] \, dt_r \quad (8)$$

where $\psi_r^{(g)}(t_r,\tau)$ denotes the recolliding electron wavepacket perturbed by the additional gate pulse as marked by the superscript (g) to be distinguished from the gate-free counterpart $\psi_r(t)$ (cf. Eq. 7). Since the gate-free quantity $\psi_r(t)$ is of interest in this discussion, the subject in this section is how to express $\tilde{\psi}_t(p,\tau)$ in terms of $\psi_r(t)$ with phase terms introduced by the gate.

First, we investigate the perturbed recolliding electron wavepacket $\psi_r^{(g)}(t_r,\tau)$ in Eq. 8 and how to link it with the gate-free electron wavepacket $\psi_r(t)$. If the gate field is sufficiently weak, i.e. $\eta = |A_g(t)|^2/|A_p(t)|^2 \ll 1$, the dipole transition and scattering amplitudes can be considered invariant, i.e., $[E_p(t) + E_g(t+\tau)]d(p + A_p(t) + A_g(t+\tau)) \approx E_p(t)g(p + A_p(t))$ and $g(p + A_p(t) + A_g(t+\tau)) \approx g(p + A_p(t))$ in the expression of the recolliding electron wavepacket (Eq. 7). In such case, the gate only modifies the phase imparted on the wavepacket between ionization and recollision:

$$\psi_r^{(g)}(t_r,\tau) \approx \int_{-\infty}^{t_r} dt' \int dp' \, g\left(p' + A_p(t_r)\right) E_p(t') d\left(p' + A_p(t')\right)$$
$$\times \exp\left[-iS\left(p', t_r, t'; A_p(t) + A_g(t+\tau)\right) - i\frac{p^2}{2}(t_r - t') + i\phi t'\right] \quad (9)$$

In view of these considerations the variation of the phase (cf. Eq. 6) can be expressed as:

$$S(p, t_2, t_1; A_p(t) + A_g(t+\tau)) = \int_{t_1}^{t_2} \left[\frac{1}{2}[p + A_p(t) + A_g(t+\tau)]^2 - \frac{1}{2}p^2\right] dt$$

$$\approx \underbrace{\int_{t_1}^{t_2} \left[\frac{1}{2}[p + A_p(t)]^2 - \frac{1}{2}p^2\right] dt}_{=S(p,t_2,t_1;A_p(t))} + \underbrace{\int_{t_1}^{t_2} [p + A_p(t)]A_g(t+\tau)\, dt}_{\equiv \Delta S(p+A_p(t),t_2,t_1;A_g(t+\tau))} \quad (10)$$

Here, the square term of $A_g(t+\tau)$ is ignored, since its contribution is negligible compared to the other terms. Eq. 10 implies that the gate field introduces an additional phase of $\Delta S\big(p + A_p(t), t_2, t_1; A_g(t+\tau)\big)$ to the gate-free case. As a result, Eq. 9 can be recast as:

$$\psi_r^{(g)}(t_r, \tau) \approx \int_{-\infty}^{t_r} dt' \int dp' g\big(p' + A_p(t_r)\big) E_p(t') d\big(p' + A_p(t')\big)$$

$$\times \exp\left[-iS\big(p', t_r, t'; A_p(t)\big) - i\frac{p'^2}{2}(t_r - t') + i\phi t'\right] \exp\left[-i\Delta S\big(p' + A_p(t), t_r, t'; A_g(t+\tau)\big)\right] \quad (11)$$

Note that if the last phase term $e^{-i\Delta S}$ was missing the expression would be identical to Eq. 7 and thus the gate-free recollision wavefunction $\psi_r(t_r)$. Hence, it would obviously be convenient to pull the $e^{-i\Delta S}$ term out of the integrals, as this would enable to express the recolliding electron wavepacket $\psi_r^{(g)}(t_r, \tau)$ in Eq. 8 via the unperturbed counterpart $\psi_r(t_r)$ and an additional phase. To further proceed with this idea, we consider two approximations.

First, following the famous saddle point approximation, the dominant contribution in the integration over the canonical momenta $p'$ is provided by the kinetic momentum $p' + A_p(t)$ that equals the kinetic momentum $p_r$ of the recolliding electron at the surface. Hence, the kinetic momentum term in $\Delta S$ may be approximated as $\Delta S(p' + A_p(t), t_r, t'; A_g(t+\tau)) \approx \Delta S(p_r, t_r, t'; A_g(t+\tau))$.

Second, since the exponent term of the additional phase $e^{-i\Delta S}$ is oscillating slowly compared to $e^{-iS}$ in the time integration over $t'$ within Eq. 11, the additional phase $\Delta S$ can be approximated by a time average $\overline{\Delta S}$ within a time window $\Delta t$:

$$\overline{\Delta S} = \frac{1}{|\Delta t|} \int_{t_r - \Delta t}^{t_r} \Delta S(p_r, t_r, t', A_g(t+\tau))\, dt' \quad (12)$$

As our main goal is to reconstruct attosecond electron wavepackets that contribute to the spectral cut-off, we choose $\Delta t$ as the time interval between ionization and recollision of the classical backscattering trajectory that results in the highest final kinetic energy. In order to evaluate the averaged phase $\overline{\Delta S}$ and therewith simplify the analytical form of Eq. 12, $\Delta S$ can be expressed as:

$$\Delta S\left(p_r, t_r, t', A_g(t+\tau)\right) = \int_{t'}^{t_r} p_r A_g(t+\tau)\, dt = -\int_{t_r}^{\infty} p_r A_g(t+\tau)\, dt + \int_{t'}^{\infty} p_r A_g(t+\tau)\, dt \quad (13)$$

By inserting Eq. 13 into Eq. 12, the effective (averaged) phase variation $\overline{\Delta S}$ can now be evaluated:

$$\overline{\Delta S} = -\int_{t_r}^{\infty} p_r A_g(t+\tau)\, dt + \int_{t_r}^{\infty} p_r \underbrace{\left[\frac{1}{\Delta t}\int_{-\Delta t}^{0} A_g(t+t'+\tau)\, dt'\right]}_{\equiv \bar{A}_g^{(b)}(t+\tau)} dt$$

$$= -\int_{t_r}^{\infty} p_r \left[A_g(t+\tau) - \bar{A}_g^{(b)}(t+\tau)\right] dt$$

$$= -\Delta S(p_r, \infty, t_r; A_g(t+\tau) - \bar{A}_g^{(b)}(t+\tau)) \quad (14)$$

where $\bar{A}_g^{(b)}(t)$ is defined as:

$$\bar{A}_g^{(b)}(t) = \frac{1}{\Delta t}\int_{-\Delta t}^{0} A_g(t+t')\, dt' \quad (15)$$

Employing the above-described approximations now enables to pull the additional phase term out of the integrations in Eq. 11 and considering the sign flip $\overline{\Delta S} \to -\overline{\Delta S}$ at the backscattering instance, the perturbed recolliding electron wavepacket can be expressed via the gate-free wavepacket and the additional averaged phase term $\psi_r^{(g)}(t_r, \tau) \approx \psi_r(t_r)\, e^{i\overline{\Delta S}}$ in Eq. 8.

We now move on to discuss how the electron wavepacket can be described at the end of the interaction in the presence of the pump and gate fields (Eq. 8). Taking the results of Eqs. 11-14 and restoring $p_r$ with the kinetic momentum $p + A_p(t)$, Eq. 8 can be rewritten as:

$$\tilde{\psi}_t(p, \tau) \propto i \int_{-\infty}^{\infty} \psi_r(t_r) \exp\left[i\frac{p^2}{2}t_r\right] \exp\left[-iS\left(p, \infty, t_r; A_p(t) + A_g(t+\tau)\right)\right]$$

$$\times \exp[-i\Delta S(p + A_p(t), \infty, t_r; A_g(t+\tau) - \bar{A}_g^{(b)}(t+\tau))]\, dt_r \quad (16)$$

Since the integration ranges for S and ΔS are identical (from $t_r$ to ∞), the two phases can be merged into a single equation ($S' = S + \Delta S$),

$$S'(p, \infty, t_r, \tau) = \int_{t_r}^{\infty} \left[\frac{1}{2}[p + A_p(t) + A_g(t+\tau)]^2 - \frac{1}{2}p^2\right] dt + \int_{t_r}^{\infty} \left(p + A_p(t)\right)\left(A_g(t+\tau) - \bar{A}_g^{(b)}(t+\tau)\right) dt$$

$$\approx \int_{t_r}^{\infty} \left[\frac{1}{2}[p + A_p(t) + \underbrace{2A_g(t+\tau) - \bar{A}_g^{(b)}(t+\tau)}_{\equiv A_{HAS}(t+\tau)}]^2 - \frac{1}{2}p^2\right] dt$$

$$= S(p, \infty, t_r; A_p(t) + A_{HAS}(t+\tau)) \qquad (17)$$

where $A_{HAS}(t)$ is hereafter referred to as *effective HAS vector potential* and reads:

$$A_{HAS}(t) = 2A_g(t) - \bar{A}_g^{(b)}(t) \qquad (18)$$

This expression of the effective HAS vector potential is compatible with the classical momentum accumulation during the excursion from the ionization to the detection, $\Delta p = -e[2A(t_r) - A(t_r - \Delta t)]$, under the rescattering condition, $\int_{t_r-\Delta t}^{t_r} A_g(t) \, dt = \Delta t \, A_g(t_r - \Delta t)$ [2,36,55]. Using Eq. 17 the terminal electron amplitude $\tilde{\psi}_t(p, \tau)$ can be expressed as:

$$\tilde{\psi}_t(p, \tau) \propto i \int_{-\infty}^{\infty} \psi_r(t_r) \exp\left[i\frac{p^2}{2}t_r\right] \exp[-iS(p, \infty, t_r; A_p(t) + A_{HAS}(t+\tau))] \, dt_r \qquad (19)$$

The above Eq. 19 implies that the gate additionally contributes to the terminal momentum of the electron wavepacket by $A_{HAS}(t_r + \tau)$ which depends on the time delay $\tau$. As described in Eq. 5, the momentum contribution $A_p(t)$ of the pump field is already incorporated in the gate-free terminal electron spectral amplitude $\tilde{\psi}_t(p)$, whose intensity is directly accessible in experiments. Therefore, it is convenient for the analysis of HAS data to express Eq. 19 with the terminal form of $\tilde{\psi}_t(p)$. Thereto, we decompose the phase in Eq. 19 via $S(p, t_2, t_1; A_p(t) + A_g(t+\tau)) \approx S(p, t_2, t_1; A_p(t)) + \Delta S(p + A_p(t), t_2, t_1; A_{HAS}(t+\tau))$ and rewrite Eq. 19 as:

$$\tilde{\psi}_t(p, \tau) \propto i \int_{-\infty}^{\infty} \psi_r(t_r) \exp\left[i\frac{p^2}{2}t_r\right] \exp\left[-iS\left(p, \infty, t_r; A_p(t)\right)\right]$$

$$\times \exp\left[-i\Delta S\left(p + A_p(t), \infty, t_r; A_{HAS}(t+\tau)\right)\right] dt_r \qquad (20)$$

In analogy to Eq. 11, if the $e^{-i\Delta S}$ term vanishes, the above equation is identical to Eq. 5, that links the recolliding electron wavepacket $\psi_r(t)$ to the terminal spectral amplitude $\tilde{\psi}_t(p)$. Here, similar to the Fourier representation of the recollision wavepacket, we define the Fourier pair of the *terminal electron wavepacket* as:

$$\tilde{\psi}_t(p) \equiv \int_{-\infty}^{\infty} dt\, \psi_t(t) \exp\left[i\frac{p^2}{2}t\right]$$
$$\psi_t(t) \equiv \int_{-\infty}^{\infty} p\, dp\, \tilde{\psi}_t(p) \exp\left[-i\frac{p^2}{2}t\right] \quad (21)$$

Note that the terminal electron wavepacket $\psi_t(t)$ is an auxiliary electron wavepacket which contains time-structure information of the recolliding electron wavepacket $\psi_r(t)$ at the recollision surface with the momenta translated by the Volkov propagation with the exponent $\exp[-iS(p,\infty,t_r;A_p(t))]$ (cf. Eqs. 1 and 5), however without the phase from space-propagation to the detection. Using the terminal electron wavepacket $\psi_t(t)$ (Eq. 21), the terminal electron spectral amplitude (Eq. 20) can be further simplified as:

$$\tilde{\psi}_t(p,\tau) \propto i \int_{-\infty}^{\infty} \psi_t(t_r) \exp\left[i\frac{p^2}{2}t_r\right] \exp[-iS(p,\infty,t_r;A_{HAS}(t+\tau))]\, dt_r \quad (22)$$

under the condition that the variation of the vector potential is weak during the time window of the recollision. The HAS spectrogram equation then reads:

$$I(p,\tau) = |\tilde{\psi}_t(p,\tau)|^2 \propto \left|\int_{-\infty}^{\infty} \psi_t(t_r) \exp\left[i\frac{p^2}{2}t_r\right] \exp[-iS(p,\infty,t_r;A_{HAS}(t+\tau))]\, dt_r\right|^2 \quad (23)$$

Eq. 23 describes a spectrogram whose reconstruction allows access to the final electron wavepacket $\psi_t(t)$ and correspondingly $\tilde{\psi}_t(p)$ as well as $A_{HAS}(t)$.

**The effective HAS vector potential $A_{HAS}(t)$.** An explicit relationship between the incident gate vector potential $A_g(t)$ and the effective HAS vector potential $A_{HAS}(t)$ can be best understood in the Fourier domain. Using the Fourier expansion, $A_g(t) = \int_{-\infty}^{\infty} d\omega\, \tilde{A}_g(\omega) e^{i\omega t}$, the effective HAS vector potential can be expressed as,

$$A_{HAS}(t) = 2\int_{-\infty}^{\infty} \tilde{A}_g(\omega)e^{i\omega t}\,d\omega - \frac{1}{\Delta t}\int_{-\Delta t}^{0}\int_{-\infty}^{\infty}\tilde{A}_g(\omega)e^{i\omega(t+t')}\,d\omega\,dt'$$

$$= \int_{-\infty}^{\infty}\tilde{A}_g(\omega)\left[2 - \frac{i}{\omega\Delta t}(e^{-i\omega\Delta t} - 1)\right]e^{i\omega t}\,d\omega$$

$$= \int_{-\infty}^{\infty}\tilde{A}_g(\omega)\tilde{g}(\omega)e^{i\omega t}\,d\omega \qquad (24)$$

where the newly introduced multiplier $\tilde{g}(\omega)$ is defined as:

$$\tilde{g}(\omega) = \left[2 - \frac{i}{\omega\Delta t}(e^{-i\omega\Delta t} - 1)\right] \qquad (25)$$

As shown in Eq. 24 the Fourier components of the effective HAS vector potential $\tilde{A}_{HAS}(\omega)$ is related to those of the incident gate vector potential $\tilde{A}_g(\omega)$ by multiplication of $\tilde{g}(\omega)$

$$\tilde{A}_{HAS}(\omega) = \tilde{A}_g(\omega)\tilde{g}(\omega) \qquad (26)$$

The gate multiplier $\tilde{g}(\omega)$ is independent from $\tilde{A}_g(\omega)$. This allows the possibility of the complete characterization of $A_g(t)$ from $A_{HAS}(t)$ imprinted in a HAS spectrogram.

To better visualize the concept of $A_{HAS}(t)$ and to verify the validity of the assumptions used in the above derivation, a semiclassical simulation of a HAS spectrogram was performed using single-cycle pulses. The photoelectron spectrum cutoff energy variation evaluated by the HAS spectrogram is compared with the effective HAS vector potential $A_{HAS}(t)$ calculated using Eq. 24 (Extended Data Fig. 6). The multiplier $\tilde{g}(\omega)$ depends on the excursion time $\Delta t$ between ionization and backscattering event of the highest energy electron. Based on the well-established recollision model, 0.685 times of central excursion period[56,57] (central period of $E(\omega)/\omega^2$), which corresponds to ~$0.84\,T_L$ was used for $\Delta t$ to evaluate $\tilde{g}(\omega)$. Here $T_L$ is the centroid period of the laser pulse. Extended Data Figs. 6b-d show that the cutoff energy variation in a HAS spectrogram excellently follows $A_{HAS}(t)$ (black curve) as calculated by the unmodified vector potential of the incident pulse (red dashed curve) regardless of the carrier-envelope phase (CEP).

# Retrieval of the vector potential $A_g(t)$ from a HAS spectrogram

The above discussion suggests that by tracing the variation of the cutoff energy in a HAS spectrogram we can gain access into $A_{HAS}(t)$ (red curves in Extended Data Fig. 7a and b). Therefore, access to the Fourier components of the effective HAS vector potential allows the characterization of the vector potential of the incident gate: $\tilde{A}_g(\omega) = \tilde{g}^{-1}(\omega)\tilde{A}_{HAS}(\omega)$ (Extended Data Fig. 7c and d). The retrieved incident vector potential $A_g(t)$ is shown in blue in Extended Data Fig. 7b.

**Identification of the absolute zero-delay in a HAS spectrogram.** Identification of the zero-delay between pump and gate pulses in a HAS spectrogram can be obtained with various methods. Here, we opted for a method that allows the absolute delay to be derived directly from the HAS spectrogram. Even though the difference between the intensities of pump and gate pulses is more than 2 orders of magnitude, discernible modulations (~ 5-10%) of the spectral amplitude of the spectrogram remain. In a HAS spectrogram, the total photoelectron yield variation can be evaluated by spectral integration at each delay point (Extended Data Fig. 7e). The absolute zero-delay point can be found as the delay point where the yield is maximally varied (vertical line, Extended Data Fig. 7e).

**Benchmarking HAS via EUV attosecond streaking.** EUV attosecond streaking provides access to the detailed field waveform of a pulse[32–34]. Because this technique of field characterization is integrated in our experimental setup it allows to benchmark HAS as a field characterization method.

Extended Data Figs. 8a and b display HAS and EUV attosecond streaking measurements, respectively. The vector potential waveform of the incident gate pulse retrieved from the cutoff analysis in HAS (red curve, Extended Data Fig. 8c) and that from EUV attosecond streaking (blue curve, Extended Data Fig. 8c) show an excellent agreement as verified by the degree of similarity of ~0.95 [44] and support the notion that the gate pulse indeed acts as a phase gate.

**The gate pulse as a phase gate**

The compact description of HAS as a spectrogram implied by Eqs. 2 and 23 assumes that the weak replica of the pump field acts as a nearly pure phase gate on the electrons released by the pump. In other words, it can modify the momentum of electrons released by the pump field but does not significantly influence the process of electron ionization. Yet, unless the ionization nonlinearities are well understood (for instance in atoms) a theoretical estimate of the required ratio between pump and gate pulses for attaining a sufficiently pure phase gate requires an experimental validation.

In order to identify safe limits within which the above condition is met, we performed HAS measurements under different gate strengths and compared the vector potential waveforms extracted from HAS to those characterized by EUV attosecond streaking. As shown in Extended Data Figs. 9a and b, the vector potential waveforms from two techniques achieve best agreement at low gate/pump intensity ratio ($\eta < 10^{-2}$). At higher intensity ratios we observe a gradually rising disagreement between the reconstructed waveforms with the two methods (Extended Data Figs. 9c and d), implying that the gate pulse does not serve any more as a weak perturbation. These measurements suggest that for the studied system HAS measurements require a gate pulse whose intensity is ~ $10^{-2}$ lower than the pump intensity.

**HAS reconstruction methodology**

At the first stage of the reconstruction of the HAS spectrogram, the terminal electron wavepacket $\psi_t(t)$ is retrieved, since its spectral intensity $|\tilde{\psi}_t(p)|^2$ can be directly obtained by a gate-free photoelectron spectrum. Therefore, the reconstruction problem is reduced to retrieval of the spectral phase.

As motivated in the main text we isolated a spectral area of interest (AOI) from 80 eV to 230 eV (Extended Data Figs. 10a and b). With this region of interest, the terminal wavepacket can be expressed as:

$$\psi_t^{(AOI)}(t) = \int_{-\infty}^{\infty} \left|\tilde{\psi}_t^{(AOI)}(\omega)\right| e^{-i\varphi(\omega)} e^{i\omega t}\, d\omega \qquad (27)$$

Here $\varphi(\omega)$ is the spectral phase of the electron wavepacket modeled as a polynomial series up to 6$^{th}$ order,

$$\varphi(\omega) = \sum_n^{N=6} D_n (\omega - \omega_c)^n \qquad (28)$$

where $D_n$ and $\omega_c$ are n$^{th}$ order dispersion and central frequency, respectively. The reconstruction is based on a least-square algorithm written in MATLAB which aims at the total minimization of the difference among the experimental (Fig. 4a and Extended Data Fig. 10a) and reconstructed spectrogram (Fig. 4b and Extended Data Fig. 10b). In order to further increase the fidelity of the reconstruction we also simultaneously fit the differential map $D(E,\tau)$ of a HAS spectrogram $I(E,\tau)$ which is defined as:

$$D(E,\tau) = \int \frac{\partial I(E,\tau)}{\partial \tau}\, d\tau \qquad (29)$$

The differential map is useful because it can eliminate the unmodulated intensity along the delay axis and allows the retrieval algorithm to reconstruct fine details of the experimental trace (Extended Data Figs. 10c-e). As an initial guess for the phase, zero phase was used. The retrieval of the terminal electron wavepacket is shown in Extended Data Figs. 10f and g.

In a next stage of the reconstruction, the recolliding electron pulse which is the key quantity in this work is evaluated by the inverse Volkov propagation of the retrieved terminal electron wavepacket as:

$$\psi_r^{(AOI)}(t) = \int_{-\infty}^{\infty} dp\, \tilde{\psi}_t^{(AOI)}(p) \exp\left[-i\frac{p^2}{2}t\right] \exp[iS(p,\infty,t;A_p(t))] \qquad (30)$$

The Volkov basis is reconstructed by precisely measuring the pump field waveform and its timing with respect to the emission. The retrieved recolliding electron pulse is shown in Fig. 4 of the main text.

**Methods references**

**Acknowledgements**

We thank J. Apportin and B. Kruse for support with FDTD simulations. This work was supported by the Deutsche Forschungsgemeinschaft (DFG, German Research Foundation) - SFB 1477 "Light-Matter Interactions at Interfaces", project number 441234705.

**Author contributions**

H.Y.K., M.G. and S.M. performed experiments and analyzed the experimental data. H.Y.K, M.G., L.S. and T.F. performed theoretical modeling and simulation. E.G. conceived and supervised the project. All authors contributed to the preparation of the manuscript.

**Competing interests**

The authors declare no competing interests.

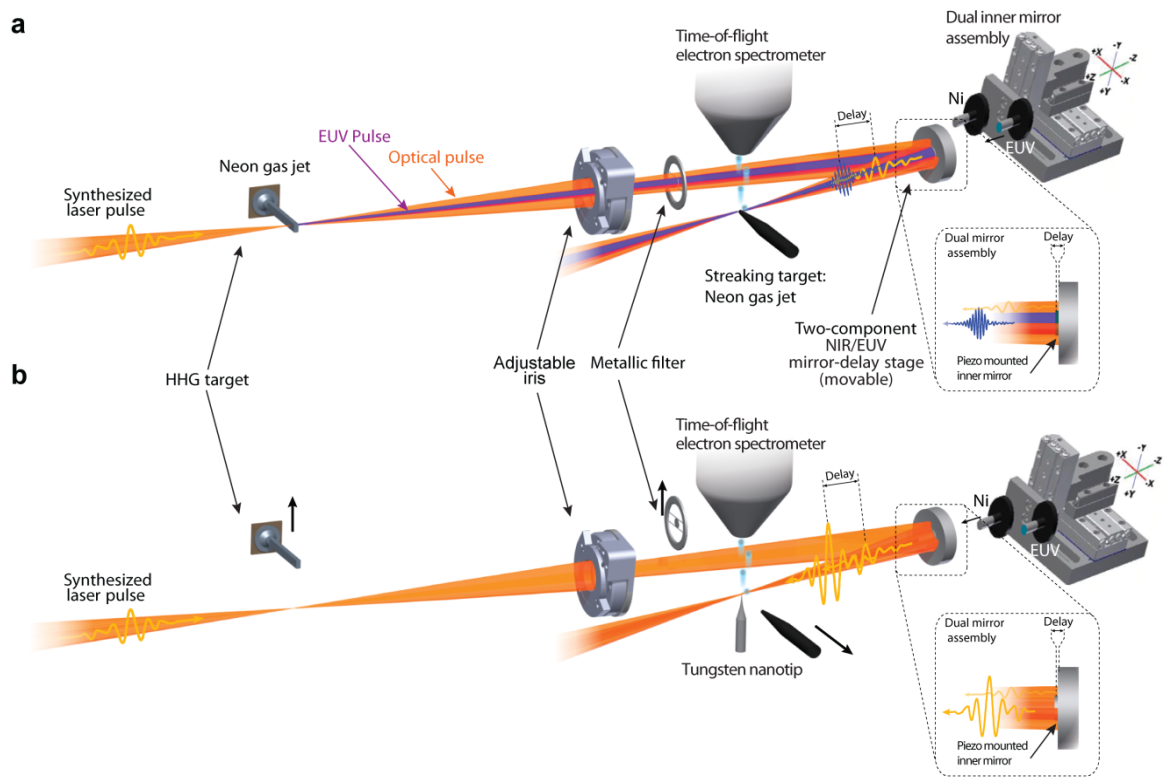

**Extended data Fig. 1 | Experimental setup.** (**a**) EUV and (**b**) Homochromic Attosecond Streaking configurations.

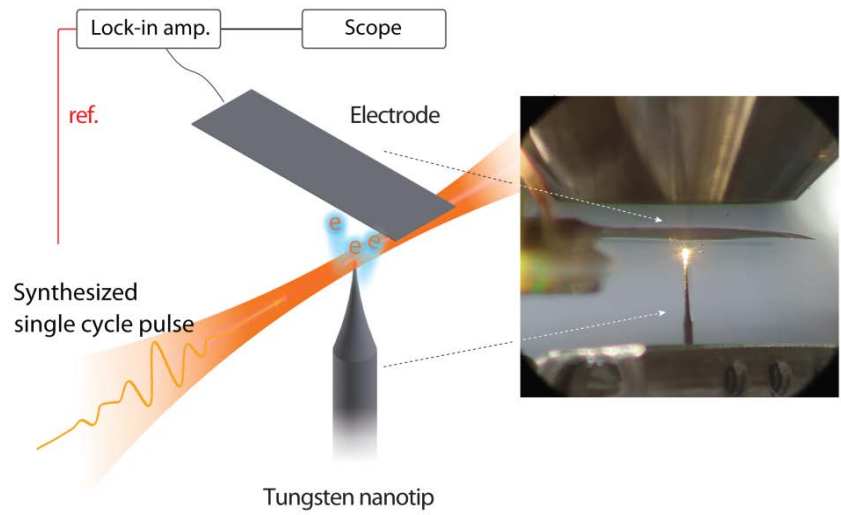

**Extended data Fig. 2 | Module for the measurement of the absolute electron yield.**

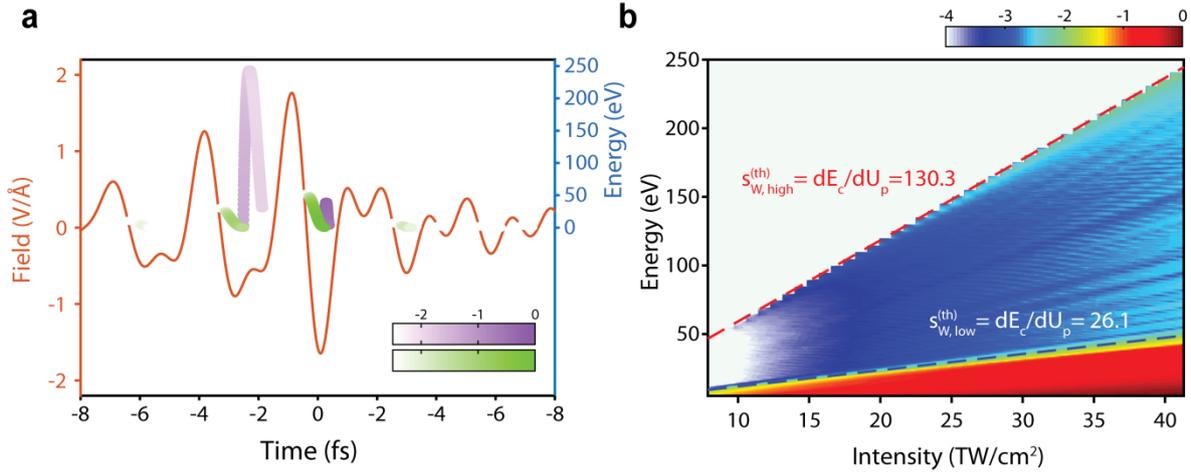

**Extended data Fig. 3 | Semiclassical simulations of the optical field emission from tungsten nanotips in the single-cycle limit. a**, Terminal electron energy as a function of the ionization time under the electric field of the incident single-cycle pulses (red curve). A peak field intensity of 41 TW/cm² was used in accordance with the experiments. Purple and green lines indicate terminal energies of backscattered and direct electrons as function of birth time, where the ionization rate is encoded on the line shading. **b**, Electron spectra plotted as a function of the peak field intensity. Red and blue dashed lines indicate the high and low cutoff energies, respectively.

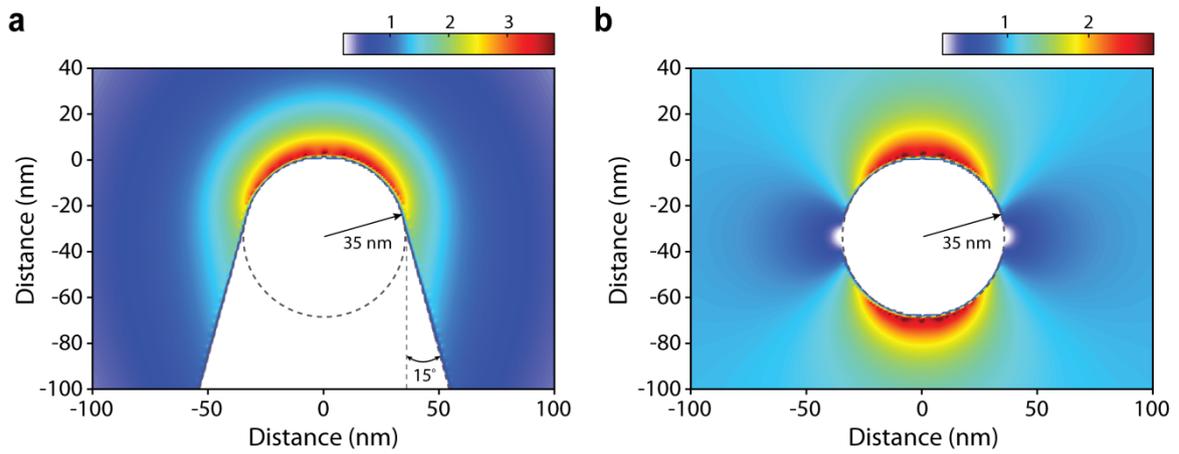

**Extended data Fig. 4 | Finite-difference time-domain (FDTD) simulations of the near field enhancement.** Evaluated spatial distribution of the field enhancement factor at (**a**) the apex of a tungsten nano-tip of an apex radius of 35nm and opening angle of 15 ° and (**b**) a similarly sized tungsten sphere.

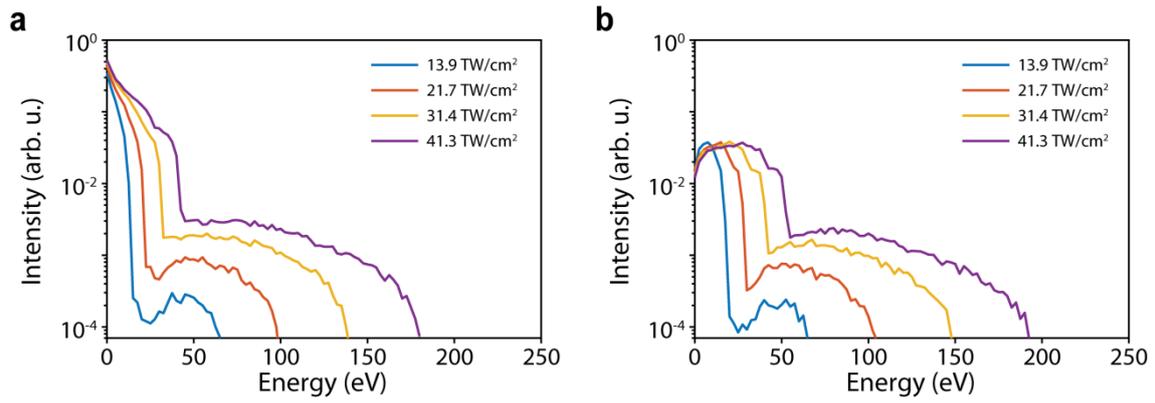

**Extended data Fig. 5 | Impact of charge interactions on the electron emission.** Electron energy spectra obtained from three-dimensional semiclassical M³C trajectory simulations for the experimental parameters (**a**) without the inclusion of charge interactions and (**b**) with the inclusion of charge interactions for different laser intensities.

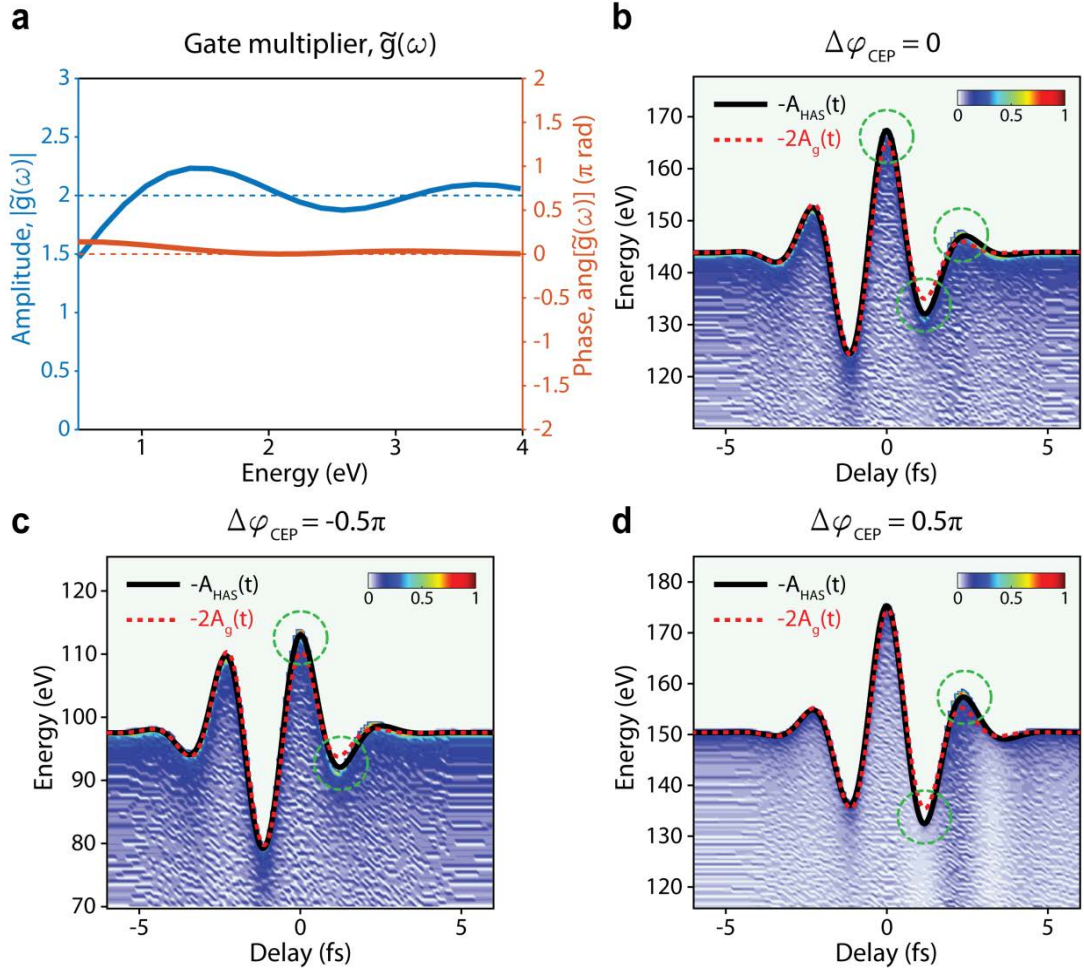

**Extended data Fig. 6 | Incident and effective HAS gate vector potentials. a,** Gate multiplier function $\tilde{g}(\omega)$ as defined in the text. **b-d,** Semiclassically simulated HAS spectrograms at different CEP settings of the driving pulse. The red dashed and black solid curves indicate waveforms of the incident gate vector potential $A_g(t)$ and effective HAS gate vector potential $A_{HAS}(t)$ calculated using Eq. 24, respectively.

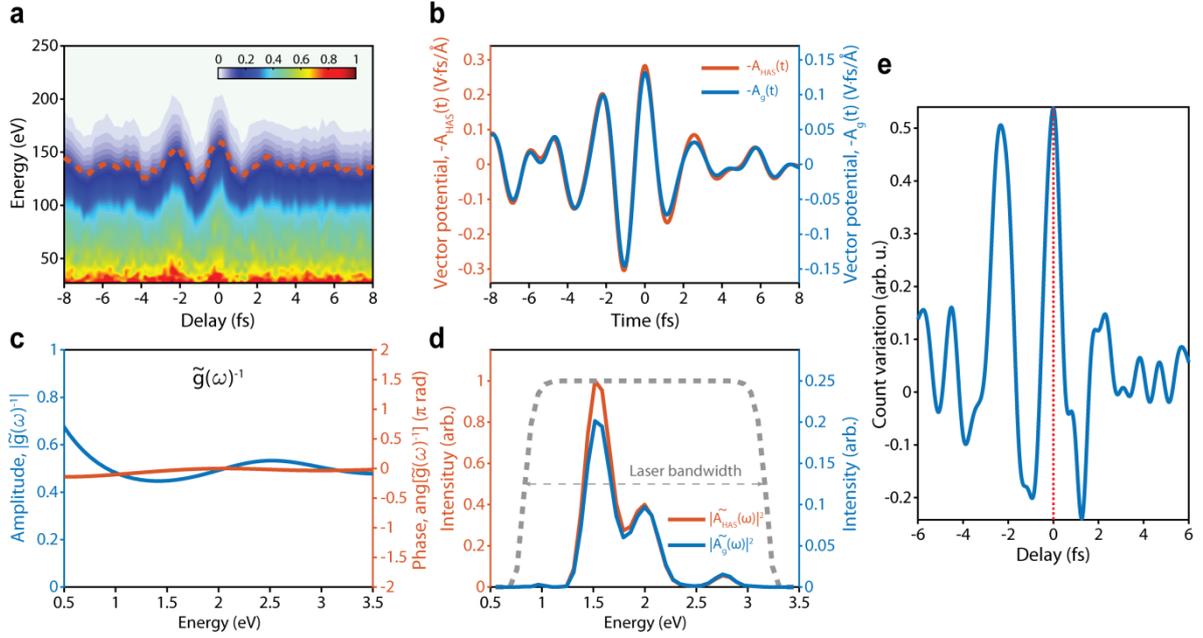

**Extended data Fig. 7 | Retrieval of the vector potential of the incident gate field. a**, Measured HAS spectrogram (Fig. 3a in main text). **b**, $A_{HAS}(t)$ (red) directly extracted via tracing the cutoff energy variation in the spectrogram and the retrieved incident vector potential $A_g(t)$ (blue). **c**, Amplitude (blue) and phase (red) of the inverse of the multiplier $\tilde{g}(\omega)^{-1}$. **d**, Spectral intensity of $A_{HAS}(t)$ (red) and incident $A_g(t)$ (blue) vector potentials. **e,** Identification of timing between gate and pump pulses in a HAS spectrogram. The variation of electron yields is evaluated by the spectral integration of the HAS spectrogram (blue curve) for each delay. The absolute zero-delay between gate and pump pulses is identified as the point at which the yield variation is maximized (red dashed line).

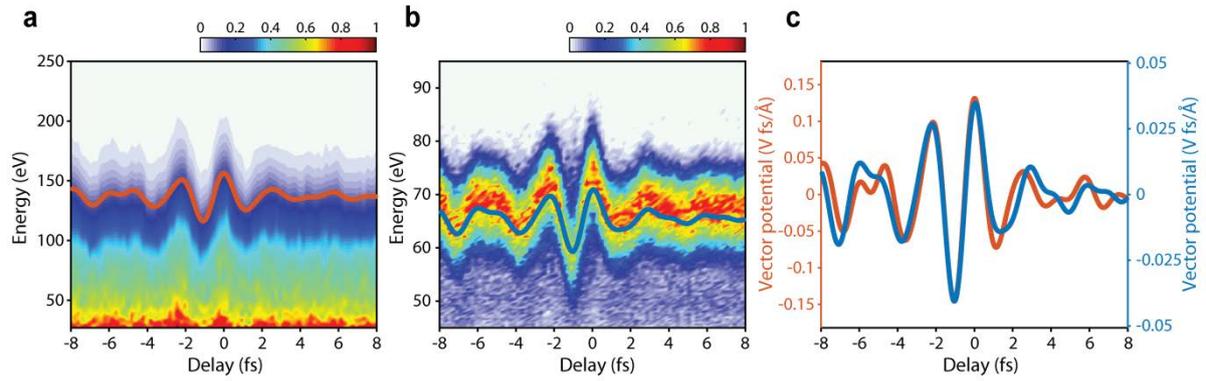

**Extended data Fig. 8 | Benchmarking of HAS via EUV attosecond streaking. a**, **b**, HAS (**a**) and EUV attosecond streaking (**b**) spectrograms under the identical driving pulse field. **c**, Comparison of the vector potential of the gate pulse evaluated from HAS (red) and EUV attosecond streaking (blue) techniques.

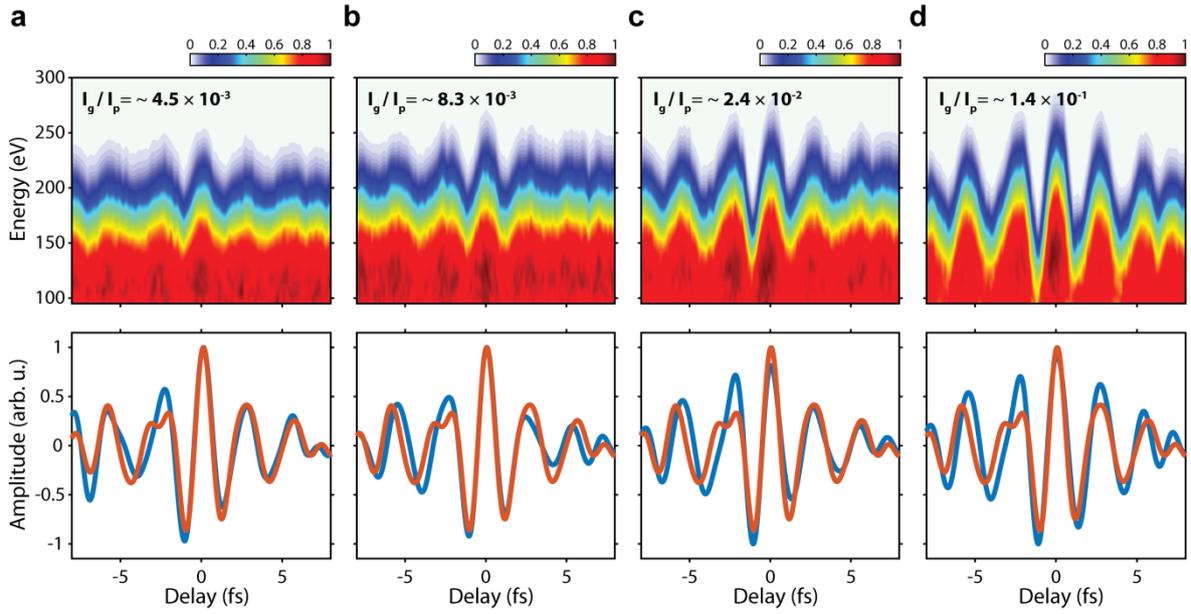

**Extended data Fig. 9 | Benchmarking HAS.** (Top panels) HAS spectrograms recorded at different gate/pump intensity ratios, $\eta \sim 4.5 \times 10^{-3}$ (**a**), $8.3 \times 10^{-3}$ (**b**), $2.4 \times 10^{-2}$ (**c**) and $1.4 \times 10^{-1}$ (**d**). The bottom panels show comparison of the vector potential waveform characterized by EUV attosecond streaking (red curve) and HAS (blue curve) for each gate/pump intensity ratio.

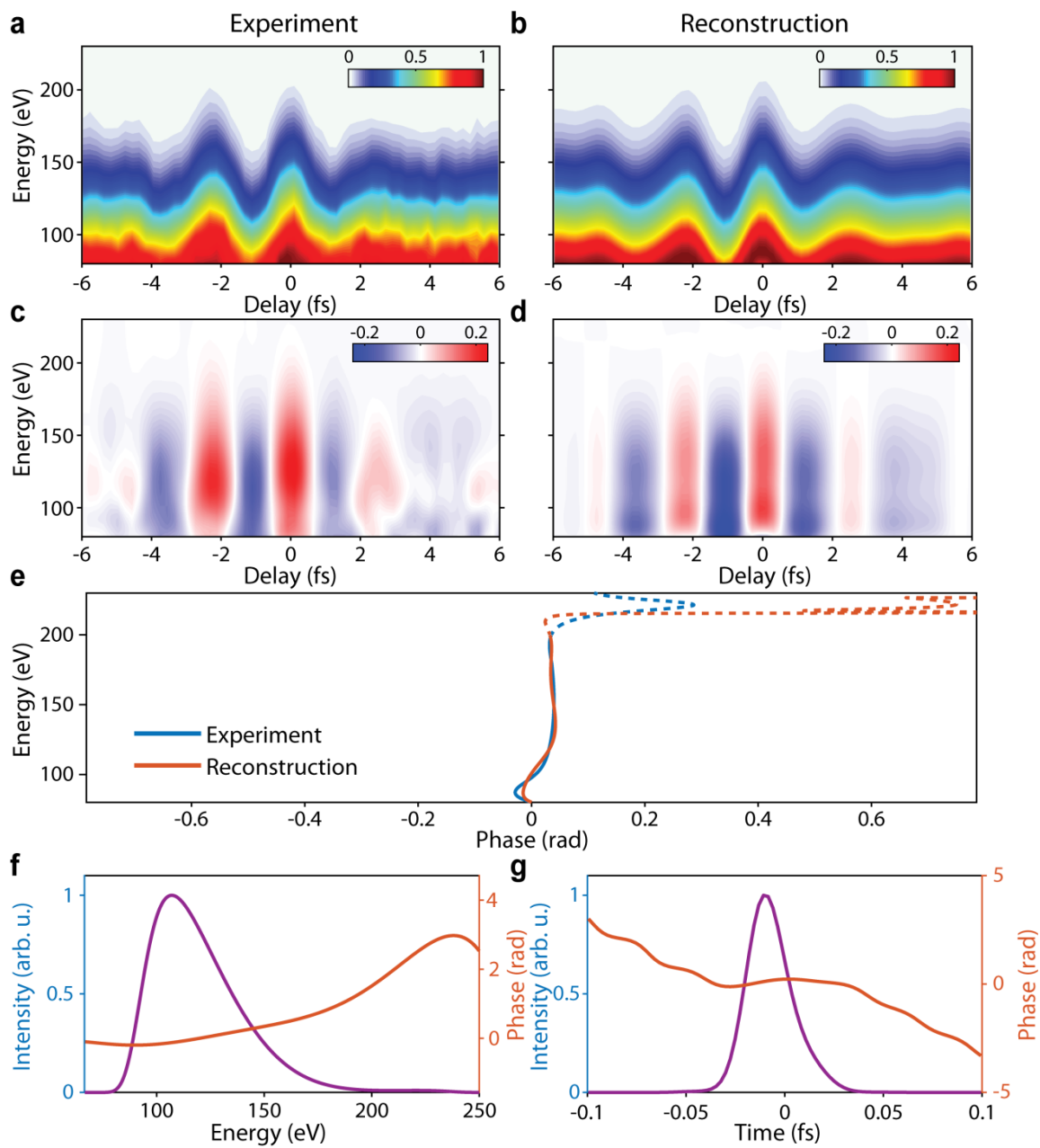

**Extended data Fig. 10 | Retrieval of the terminal electron wavepacket. a**, **b**, Measured (**a**) and reconstructed (**b**) HAS spectrograms in the energy range from 80 to 230 eV. **c**, **d**, Differential maps of measured (**c**) and reconstructed (**d**) spectrogram. **e**, The phase of the intensity shift in measured (blue) and reconstructed (red) spectrograms. **f**, **g**, Retrieved terminal electron wavepacket in spectral (**f**) and temporal (**g**) domains. The purple and red curves denote the intensity and phase, respectively.